\documentclass[letterpaper,12pt]{article}

\usepackage{cite}

\usepackage{fancyheadings}
\usepackage[dvips]{graphicx}
\usepackage{graphpap}
\usepackage{ifthen}
\usepackage{enumerate}
\usepackage{cite}
\usepackage{amssymb}
\usepackage[mathscr]{eucal}
\usepackage[errorshow]{tracefnt}
\usepackage{fancybox}
\usepackage{amsfonts}
\usepackage{amsmath}
\usepackage{amssymb}
\usepackage{amsthm}
\usepackage{eucal}
\usepackage{eufrak}

\newcommand{\cM}{\ensuremath{\mathcal{M}}}
\newcommand{\cN}{\ensuremath{\mathcal{N}}}
\newcommand{\cP}{\ensuremath{\mathcal{P}}}
\newcommand{\cQ}{\ensuremath{\mathcal{Q}}}

\newcommand\nn{\nonumber}

\newcommand{\bor}{U}

\newcommand{\beq}{\begin{equation}}
\newcommand{\beqn}{\begin{equation*}}
\newcommand{\eeq}{\end{equation}}
\newcommand{\eeqn}{\end{equation*}}
\newcommand{\beqa}{\begin{eqnarray}}
\newcommand{\beqan}{\begin{eqnarray*}}
\newcommand{\eeqa}{\end{eqnarray}}
\newcommand{\eeqan}{\end{eqnarray*}}
\newcommand{\bdm}{\begin{displaymath}}
\newcommand{\edm}{\end{displaymath}}

\newcommand{\la}{\langle}
\newcommand{\ra}{\rangle}

\def\der'{\mathfrak{der}'\,}
\def\der{\mathfrak{der}\,}
\def\str'{\mathfrak{str}'\,}
\def\str{\mathfrak{str}\,}

\def\g{\gamma}

\def\g{\mathfrak{g}}

\def\sl{\mathfrak{sl}}

\newcommand{\al}{\alpha}
\newcommand{\be}{\beta}

\newcommand{\de}{\delta}

\newcommand{\dlb}{\ensuremath{[\![}}
\newcommand{\drb}{\ensuremath{]\!]}}

\RequirePackage{hyperref}

\numberwithin{equation}{section}

\begin{document}

\vskip-10pt
\hfill {\tt 1203.5107v2}\\
\vskip-10pt
\hfill {\tt \today}\\
\vskip-10pt
\hfill {\tt ULB-TH/12-05}\\

\pagestyle{empty}

\vspace*{2.5cm}

\noindent
\begin{center}
{\LARGE {\sf \textbf{ 
Borcherds and Kac-Moody extensions
}}}\\
\vspace{.2cm}
{\LARGE {\sf \textbf{
of simple finite-dimensional Lie algebras}}}\\
\vspace{.3cm}

\renewcommand{\thefootnote}{\fnsymbol{footnote}}

\vskip 1truecm
\noindent
{\large {\sf \textbf{Jakob Palmkvist}\footnote{\it Also affiliated to: Department of Fundamental Physics,
  Chalmers University of Technology,\\ SE-412 96 G\"oteborg, Sweden}}}\\
\vskip 1truecm
        {\it 
        {Physique Th\'eorique et Math\'ematique\\
  Universit\'e Libre de Bruxelles \& International Solvay Institutes\\
  Boulevard du Triomphe, Campus Plaine, ULB-CP 231,\\BE-1050 Bruxelles, Belgium}\\[3mm]}
        {\tt jakob.palmkvist@ulb.ac.be} \\
\end{center}

\vskip 1cm

\centerline{\sf \textbf{
Abstract}}
\vskip .2cm

We study the Borcherds superalgebra obtained by adding an odd (fermionic) null root to the set of simple roots of a simple finite-dimensional Lie algebra. We compare it to the Kac-Moody algebra obtained by replacing the odd null root by an ordinary simple root, and then adding more simple roots, such that each node that we add to the Dynkin diagram is connected to the previous one with a single line. This generalizes the situation in maximal supergravity, where the $E_n$ symmetry algebra can be extended to either a Borcherds superalgebra or to the Kac-Moody algebra $E_{11}$, and both extensions 
can be used to derive the spectrum of $p$-form potentials in the theory. We show that also in the general case, the Borcherds and Kac-Moody extensions lead to the same `$p$-form spectrum' of representations of the simple finite-dimensional Lie algebra.

\newpage

\pagestyle{plain}

\section{Introduction}

Maximal supergravity in $D$ dimensions contains $p$-form potentials that transform in representations of a 
global symmetry group. Including also the non-dynamical $(D-1)$- and $D$-forms that are possible to add to the theory, 
all these representations can be derived from the infinite-dimensional Kac-Moody algebra $E_{11}$ 
\cite{West:2001as, Schnakenburg:2001ya, West:2004st,West:2005gu,Riccioni:2007au,Bergshoeff:2007qi}. 
Considering $E_{11}$ as the `very extension'
of (the split real form of) the exceptional Lie algebra $E_8$,
the corresponding derivation also works for half-maximal supergravity theories, 
where the role of $E_8$ is played by $B_7$, $B_8$ or $D_8$
\cite{Bergshoeff:2007vb}.
The spectrum of $p$-form representations for maximal supergravity in $D$ dimensions can alternatively be derived from a Borcherds algebra which depends on $D$ \cite{HenryLabordere:2002dk,HenryLabordere:2003rd,Henneaux:2010ys,Greitz:2011da,Greitz:2011vh}. 
The fact that these Borcherds algebras lead to the same $p$-form spectrum as
$E_{11}$ was explained in \cite{Henneaux:2010ys}, and an alternative explanation was given in \cite{Palmkvist:2011vz}.
Since Borcherds algebras arise from Bianchi identities in half-maximal supergravity \cite{Greitz:2012} as well as in maximal theories
\cite{Greitz:2011da,Greitz:2011vh}
it raises the question whether also these Borcherds algebras
lead to the same $p$-form spectra as the
corresponding very extended Kac-Moody algebras. The present paper gives an affirmative answer to that question, 
by generalizing the result in \cite{Palmkvist:2011vz}.

The Borcherds algebra associated to maximal supergravity in
$3 \leq D \leq 7$ dimensions, with a 
Lie algebra $\g$ corresponding to the global symmetry group,
can be constructed from $\g$ by adding an extra simple root in a certain way 
(or equivalently, an extra node to the Dynkin diagram of $\g$).
It is in fact not a Lie algebra but a Lie {\it superalgebra}, where the eigenvectors corresponding to the extra simple root are odd 
elements, and furthermore the eigenvalues are zero. 
If we instead add $N \geq 1$ ordinary simple roots
(such that each node that we add to the Dynkin diagram is connected to the previous one with a single line), 
then we obtain a Kac-Moody algebra, which for $N=D$ is the 
very extended Kac-Moody algebra $E_{11}$.
Up to level $p=N$, the level decomposition of this Kac-Moody algebra under $\g \oplus \sl_N$, 
restricted to antisymmetric $\sl_N$ tensors, gives the same `$p$-form spectrum' of
$\g$ representations as the level decomposition of the Borcherds algebra under $\g$.
We will show that the corresponding result holds
for any such 
Borcherds and Kac-Moody 
extensions of a simple finite-dimensional Lie algebra $\g$. 
More precisely, for {\it any} such $\g$,
{\it any} way of adding the first extra node (the only extra node for the Borcherds algebra), 
and {\it any} total number $N$ of extra nodes for the Kac-Moody algebra, the two algebras lead to the same $p$-form spectrum 
(up to level $p=N$).
The fact that $N$ can be larger than the spacetime dimension $D$ is important for applications to
the superspace approach that has been employed recently in 
\cite{Greitz:2011da,Greitz:2011vh,Greitz:2012},
since $p$-form superfields with $p > D$ need not be zero.

In this paper we denote the
Borcherds and Kac-Moody extensions by 
$U$ and $W$, and we let $V$ be an intermediate Kac-Moody algebra. 
The algebras $U$, $V$ and $W$ are described in
section 2, 3 and 4, respectively, and in the end we show that $W$ gives the same $p$-form spectrum as $U$.
The reader who finds the paper difficult to follow is invited to read \cite{Palmkvist:2011vz} first, where
$U$, $V$ and $W$
corresponds $U_{n+1}$, $E_{n+1}$ and $E_{11}$.

\section{The Borcherds algebra $U$}

As indicated in the introduction, our definition of Borcherds algebras include also the generalization \cite{Ray95} of the original Borcherds algebras \cite{Borcherds88} to superalgebras. However, we will here only define a very special case of such superalgebras, and refer to \cite{Ray06}
for the full definition. As noted in \cite{Henneaux:2010ys}, footnote 8, there is an error 
in the definition in \cite{Ray06}, but this has no importance for the special cases we consider here.

A Borcherds algebra is given by a (generalized) Cartan matrix $a_{IJ}$, which is a non-degenerate symmetric real matrix, where the rows and columns are labelled by some index set. This set can in general be infinite, but here we restrict it to be finite and write 
$I,J,\ldots=0,1,\ldots,r$ for some $r$. For each value $I$ of the 
indices we associate two Chevalley generators $e_I$ and $f_I$ which are both either 
odd (fermionic) or even (bosonic) elements of the Borcherds algebra. 
We assume $e_0$ and $f_0$ to be odd, and use the indices $i,j,\ldots=1,2,\ldots,r$
for the even generators.
Furthermore, we assume that $a_{00}=0$ and $a_{ii}>0$. With these restrictions, the conditions that define $a_{IJ}$ to be the Cartan matrix of a Borcherds algebra are
\begin{align}
I\neq J &\Rightarrow a_{IJ}\leq 0, 
& 2\frac{a_{iJ}}{a_{ii}} &\in \mathbb{Z}. \label{cartanmatrixdef}
\end{align}
Note that this matrix is symmetric, unlike general Cartan matrices of Kac-Moody algebras
with the standard definition (see for example \cite{Kac,Fuchs}).
However, we can `de-symmetrize' $a_{IJ}$ and define an in general
non-symmetric matrix $A_{IJ}$ by 
\begin{align}
A_{iJ}&=2\frac{a_{iJ}}{a_{ii}}, & A_{0i}&=a_{0i}, & A_{00}=a_{00}=0.
\end{align}
Any multiple of $a_{IJ}$ gives the same Borcherds algebra as $a_{IJ}$. Together with the second condition in (\ref{cartanmatrixdef}), this implies that we can assume all the diagonal entries $a_{ii}$ to be even integers. It then follows from the same condition that all the entries in $a_{IJ}$ are integers, in particular $a_{0i}$. We conclude that $A_{IJ}$ is an integer-valued matrix, with $A_{00}=0$ and $A_{ii}=2$. The off-diagonal entries are non-positive integers, in general with $A_{IJ} \neq A_{JI}$, but if $A_{IJ}=0$, then 
$A_{JI}=0$ as well. Thus $A_{ij}$ satisfies the definition of a Cartan matrix of a Kac-Moody algebra, and, as a last restriction, we require this Kac-Moody algebra to be finite, that is, a simple finite-dimensional 
Lie algebra $\g$.

The Borcherds algebra $\bor$ associated to
$a_{IJ}$ (or $A_{IJ}$) is now defined as the Lie superalgebra
generated by the Chevalley generators $e_I,f_I$ and $h_I=\dlb e_I,f_I \drb$ modulo the relations
\begin{align}
\dlb h_I,e_J \drb &=A_{IJ}e_J, & 
\dlb h_I,f_J\drb &=-A_{IJ}f_J, & \dlb e_I,f_J\drb&=\delta_{IJ}h_J,\nn 
\end{align}
\begin{align} \label{serre-rel}
\dlb e_0,e_0\drb=\dlb f_0,f_0\drb=
(\text{ad }e_i)^{1-A_{iJ}}(e_J)=(\text{ad }f_i)^{1-A_{iJ}}(f_J)=0,
\end{align}
where $i \neq J$, and $\dlb x,y \drb$ denotes the supercommutator of two elements $x$ and $y$. This is a symmetric anticommutator
$\dlb x,y \drb \equiv \{x,y\}=\{y,x\}$
if both $x$ and $y$ are odd elements, and an ordinary antisymmetric commutator
$\dlb x,y \drb \equiv [x,y]=-[y,x]$
if at least one of the elements is even.

The Borcherds algebra $\bor$ has a bilinear form, which we write as $\la x| y\ra$ for two elements
$x$ and $y$,
and define by
\begin{align} \label{def-innerprod}
\la h_I | h_J \ra &= A_{IJ}, &\la e_I | f_J \ra &= \delta_{IJ}, & \la e_I | e_J \ra=\la f_I | f_J \ra=
\la h_I | e_J \ra=\la h_I | f_J \ra=0.
\end{align}
The definition can then be extended to the full algebra 
${\bor}$ in such a way that the bilinear form is invariant
\begin{align}
\la\dlb x,y \drb | z\ra &= \la x | \dlb y , z \drb \ra,
\end{align}
and supersymmetric, which means that $\la x|y\ra = -\la y|x\ra$ if both elements are odd, and $\la x|y\ra = \la y|x\ra$
if at least one of them is even.

The odd generators $e_0$ and $f_0$ give rise to a $\mathbb{Z}$-grading of $\bor$
which is consistent with the $\mathbb{Z}_2$-grading that 
$\bor$ naturally is equipped with as a superalgebra. This means that it can be written as a direct sum of subspaces 
$\bor_{p}$ for all integers $p$, such that 
\begin{align}
\dlb \bor_{p},\bor_{q} \drb \subseteq \bor_{p+q}, \label{gradering}
\end{align}
where $\bor_{p}$ consists of odd elements if $p$ is odd, and of even elements if $p$ is even.
(These subspaces should not be confused with $U_{n+1}$ in \cite{Palmkvist:2011vz}, which is simply $U$ here.) 
Among the Chevalley generators $e_0$ belongs to $\bor_{-1}$, whereas $f_0$ belongs to 
$\bor_{1}$, and all the others belong to $\bor_{0}$. 

It follows from the grading (\ref{gradering}) that each subspace $\bor_p$ constitute a representation ${\bf r}_p$ 
of $\g$ (called ${\bf s}_p$ in \cite{Palmkvist:2011vz}). 
One can easily see that there is an isomorphism between the subspaces $\bor_1$ and $U_{-1}$, such that elements mapped to each other have eigenvalues with opposite signs under the adjoint action of $h_i$, and therefore the representations ${\bf r}_1$ and 
${\bf r}_{-1}$ are conjugate to each other.
Accordingly, we introduce indices 
\begin{align}
\cM,\cN,\ldots = 1, 2, \ldots, \text{dim }{\bf r}_1,
\end{align}
and write the basis elements of $\bor_{-1}$ and $\bor_{1}$ as $E_\cM$ and $F^\cN$, respectively, chosen such that 
$\la E_\cM | F^\cN \ra=\de_\cM{}^\cN$.
For $p \geq 2$ the subspace ${\bor}_{-p}$ is then spanned by the elements
\begin{align} \label{e-uttryck}
E_{\cM_1\cdots\cM_p} \equiv \dlb E_{\cM_1},\dlb E_{\cM_2},\ldots, \dlb E_{\cM_{p-1}},E_{\cM_p} \drb \cdots \drb\drb
\end{align} 
and ${\bor}_{p}$ by the elements
\begin{align} \label{f-uttryck}
F^{\cN_1\cdots\cN_p} \equiv \dlb F^{\cN_1},\dlb F^{\cN_2},\ldots, \dlb F^{\cN_{p-1}},F^{\cN_p} \drb \cdots \drb\drb.
\end{align}
As explained in \cite{Palmkvist:2011vz}, each representation ${\bf r}_p$ is determined by the lower (or upper)
indices in the tensor
\begin{align} \label{gen-structconst}
f_{\cN_1\cdots\cN_{p}}{}^{\cP_1\cdots\cP_{p}} &= 
\la E_{\cN_1\cdots\cN_{p}} | F^{\cP_1\cdots\cP_{p}}\ra,
\end{align}
and all such tensors can be computed recursively, starting from the constants
\begin{align}
f_\cM{}^\cN{}_\cP{}^\cQ=\la [\{E_\cM,F^\cN\},E_\cP] | F^\cQ \ra,
\end{align}
which are 
the structure constants of $U_{-1}$ considered as a (generalized Jordan) triple system with the triple product $[\{E_\cM,F^\cN\},E_\cP]$. To find the recursion formula,
we first use the Jacobi identity to compute
\begin{align} 
\dlb F^{\cN}, E_{\cM_1\cdots\cM_{p}}\drb&=
\dlb \{  F^{\cN}, E_{\cM_1} \}, E_{\cM_2 \cdots \cM_p} \drb\nn\\
&\quad\, -  \dlb E_{\cM_1} , \dlb F^{\cN} , E_{\cM_2 \cdots \cM_p} \drb \drb \nn\\
&=
\dlb \{  F^{\cN}, E_{\cM_1} \}, E_{\cM_2 \cdots \cM_p} \drb\nn\\
&\quad\, -  \dlb E_{\cM_1} , \dlb \{ F^{\cN} , E_{\cM_2}\} , E_{\cM_3 \cdots \cM_p} \drb \drb \nn\\
&\quad\, +  \dlb E_{\cM_1} , \dlb  E_{\cM_2} , \dlb  F^{\cN} , E_{\cM_3 \cdots \cM_p} \drb \drb \drb \nn\\
&=
\dlb \{  F^{\cN}, E_{\cM_1} \}, E_{\cM_2 \cdots \cM_p} \drb\nn\\
&\quad\, -  \dlb E_{\cM_1} , \dlb \{ F^{\cN} , E_{\cM_2}\} , E_{\cM_3 \cdots \cM_p} \drb \drb \nn\\
&\quad\, +  \dlb E_{\cM_1} , \dlb  E_{\cM_2} , \dlb  \{ F^{\cN} , E_{\cM_3}\}, E_{\cM_4 \cdots \cM_p} \drb \drb \drb \nn\\
&\qquad\cdots\nn\\
&\quad\, +(-1)^{p+1}  \dlb E_{\cM_1} , \dlb  E_{\cM_2} , \ldots , \dlb E_{\cM_{p-1}}, \{F^{\cN} ,  E_{\cM_p}\} 
\drb \cdots \drb \drb\nn\\
&= \sum_{i=1}^{p-1} \sum_{j=i+1}^p (-1)^{i+1} f_{\cM_i}{}^\cN{}_{\cM_j}{}^\cP
E_{\cM_1 \cdots \cM_{i-1}\cM_{i+1}\cdots \cM_{j-1}\cP\cM_{j+1}\cdots \cM_p}\nn\\
&\qquad\quad\quad+(-1)^{p} f_{\cM_p}{}^\cN{}_{\cM_{p-1}}{}^\cP
E_{\cM_1 \cdots \cM_{p-2}\cP}, \label{F1Ep}
\end{align}
and then, using the invariance of the bilinear form, we obtain
\begin{align}
f_{\cM_1\cdots\cM_{p}}{}^{\cN_1\cdots\cN_{p}}&=
\la E_{\cM_1\cdots\cM_{p}} | F^{\cN_1\cdots\cN_{p}}\ra\nn\\
&=(-1)^{p+1} \la \dlb F^{\cN_1}, E_{\cM_1\cdots\cM_{p}}\drb | F^{\cN_2\cdots\cN_{p}}\ra\nn\\
&=\sum_{i=1}^{p-1} \sum_{j=i+1}^p (-1)^{i+p}
f_{\cM_i}{}^{\cN_1}{}_{\cM_j}{}^\cP
f_{\cM_1 \cdots \cM_{i-1}\cM_{i+1}\cdots \cM_{j-1}\cP\cM_{j+1}\cdots \cM_p}{}^{\cN_2\cdots\cN_p}\nn\\
&\qquad\qquad\qquad-f_{\cM_p}{}^{\cN_1}{}_{\cM_{p-1}}{}^\cP
f_{\cM_1 \cdots \cM_{p-2}\cP}{}^{\cN_2\cdots\cN_p}. \label{borcherdsutrakning}
\end{align}

The subspace $\bor_{0}$ is spanned by $\g$ and $h_0$. 
Since $U_0$ is a finite-dimensional representation of $\g$ it must be fully reducible, and since its dimension is $(\text{dim }\g+1)$
it must (as a Lie algebra) be the direct sum of $\g$ and a one-dimensional abelian subalgebra, spanned by an element $c$.
It then follows from the invariance of the bilinear form that the commutation relations between the elements in $U_{0}$ and 
$U_{\pm 1}$ are
\begin{align}
\{E_\cM,\,F^\cN\}&=(t_\alpha)_\cM{}^\cN t^\alpha + \de_\cM{}^\cN c, &
[t^\alpha,c]&=0,\nn
\end{align}
\begin{align}
[t^\alpha,E_\cM]&=(t^\alpha)_\cM{}^\cN E_\cN, & [c,E_\cM]&=\la c | c \ra E_\cM,\nn\\
[t^\alpha,F^\cN]&=-(t^\alpha)_\cM{}^\cN F^\cM, & [c,F^\cN]&=- \la c | c \ra F^\cN, \label{borcherds-comm-rel}
\end{align}
where $t^\al$ are the basis elements of $\g$, and $(t_\alpha)_\cM{}^\cN$ are the components of $t_\alpha$ in the representation 
${\bf r}_1$. The adjoint index $\alpha$ has been lowered with the restriction of the invariant bilinear form to $\g$
(the Killing form), so that 
$\la t^\al | t_\be \ra=\de^\al{}_\be$, and the normalization of $c$ has been fixed by the first equation in 
(\ref{borcherds-comm-rel}) as we will see in the next section.
Thus we end up with the expression
\begin{align} \label{lilla-f}
f_\cM{}^\cN{}_\cP{}^\cQ=\la [\{E_\cM,F^\cN\},E_\cP] | F^\cQ \ra = 
(t_\alpha)_\cM{}^\cN(t^\alpha)_\cP{}^\cQ+\la c | c \ra\,\de_\cM{}^\cN\de_\cP{}^\cQ
\end{align}
for the structure constants $f_\cM{}^\cN{}_\cP{}^\cQ$, which can then be inserted in (\ref{borcherdsutrakning}).

\section{The Kac-Moody algebra $V$}

Let $B_{IJ}$ be the matrix obtained from $A_{IJ}$ by replacing the entry $A_{00}=0$ by $B_{00}=2$. Thus we have
\begin{align}
B_{00}&=2, & B_{Ii}&=A_{Ii}, & B_{iI}&=A_{iI}.
\end{align}
We then define the Kac-Moody algebra $V$ associated to the Cartan matrix $B_{IJ}$
as the Lie algebra generated by $e_I$, $f_I$ and $h_I=[e_I,f_I]$
modulo the relations (\ref{serre-rel}), but now with $A_{IJ}$ replaced by $B_{IJ}$, and all Chevalley generators being even elements, so that the supercommutators are ordinary antisymmetric commutators. The relations corresponding to
(\ref{def-innerprod}) define a bilinear form on $V$ which is invariant and, unlike the one on $U$, fully symmetric. We write it as 
$( x| y )$ for two elements
$x$ and $y$ to distinguish it from the invariant bilinear form on $U$. Note that we have $(h_0|h_0)=2$, whereas $\la h_0| h_0 \ra =0$.

In the same way as for $U$, the generators $e_0$ and $f_0$ give rise to a $\mathbb{Z}$-grading of $V$, where each subspace 
$V_p$ constitutes a representation ${\bf s}_p$ of $\g$.
The difference between $A_{IJ}$ and $B_{IJ}$ does not affect the commutation relations between $\g$ and $e_I$ or $f_I$, and therefore we have ${\bf s}_{\pm 1}={\bf r}_{\pm 1}$. Furthermore, 
the Lie algebra 
$V_0$ is, in the same way as $U_0$, the direct sum of $\g$ and a one-dimensional abelian subalgebra spanned by an element $d$.
Using the same notation for the basis elements of $V_{\pm 1}$ as for $U_{\pm 1}$, the commutation relations
between the elements in $V_{0}$ and 
$V_{\pm 1}$ are then
\begin{align}
[E_\cM,\,F^\cN]&=(t_\alpha)_\cM{}^\cN t^\alpha + \de_\cM{}^\cN d, &
[t^\alpha,d]&=0,\nn
\end{align}
\begin{align}
[t^\alpha,E_\cM]&=(t^\alpha)_\cM{}^\cN E_\cN, & [d,E_\cM]&=( d | d ) E_\cM,\nn\\
[t^\alpha,F^\cN]&=-(t^\alpha)_\cM{}^\cN F^\cM, & [d,F^\cN]&=- ( d | d ) F^\cN. \label{km-comm-rel}
\end{align}
Let us compare $d$ in $V_0$ with the corresponding element $c$ in $U_0$. 
From the invariance of the bilinear form it follows that $c$ and $d$ are determined up to normalization by the conditions 
$\la c | \g \ra=0$ and $(d|\g)=0$, respectively. This implies in turn that both $c$ and $d$ are linear combinations 
\begin{align}
c&=c_0 h_0 + c_1 h_1 + \cdots + c_r h_r,\nn\\
d&=d_0 h_0 + d_1 h_1 + \cdots + d_r h_r
\end{align}
(identifying the Chevalley generators of 
$U$ and $V$ with each other). The first equations in (\ref{borcherds-comm-rel})
and (\ref{km-comm-rel}) fix the coefficients $c_0$ and $d_0$
to $c_0=d_0=1$. Furthermore, the conditions $\la c | \g \ra=0$ and $(d|\g)=0$ 
do not involve $A_{00}$ or $B_{00}$, which are the only entries that differ between $A_{IJ}$ or $B_{IJ}$, so they are in fact equivalent, and we conclude that $c=d$.
Now we have
\begin{align}
[c,e_0] &= c_0[h_0, e_0] + c_1[h_1,e_0] + \cdots + c_r[h_r,e_0] \nn\\
&= (c_0 A_{00}+c_1 A_{10} + \cdots + c_r A_{r0})e_0\nn\\
&= (c_1 A_{10} + \cdots + c_r A_{r0})e_0
\label{g-e0-comm}
\end{align}
in $U$, and 
\begin{align}
[d,e_0] &= d_0[h_0, e_0] + d_1[h_1,e_0] + \cdots + d_r[h_r,e_0] \nn\\
&= (d_0 B_{00}+d_1 B_{10} + \cdots + d_r B_{r0})e_0\nn\\
&= (2+c_1 A_{10} + \cdots + c_r A_{r0})e_0 
\label{g-e0-comm2}
\end{align}
in $V$. On the other hand, from (\ref{borcherds-comm-rel})
and (\ref{km-comm-rel}) we have $[c,e_0] = \la c | c \ra e_0$ in $U$, and $[d,e_0] = ( d | d ) e_0$ in $V$,
so we conclude that $( d | d )=\la c | c \ra +2$.
It follows that the structure constants of $V_{-1}$ considered as a triple system are
\begin{align}
g_\cM{}^\cN{}_\cP{}^\cQ=( [[E_\cM,F^\cN],E_\cP] | F^\cQ ) &= (t_\alpha)_\cM{}^\cN(t^\alpha)_\cP{}^\cQ+
(\la c | c \ra+2)\de_\cM{}^\cN\de_\cP{}^\cQ\nn\\
&=f_\cM{}^\cN{}_\cP{}^\cQ + 2 \de_\cM{}^\cN\de_\cP{}^\cQ.
\end{align}

\section{The extended Kac-Moody algebra $W$}
Let $C$ be the matrix obtained from $B$ by adding $N-1$ more rows and columns,
labelled by $m,n,\ldots=-N+1, -N+2, \ldots,-1$, so that
\begin{align}
C_{IJ}&=B_{IJ}, & C_{mI}&=C_{Im}=0,
\end{align}
and $C_{mn}$ is the well known Cartan matrix of $A_{N-1}=\sl_N$. Let $W$ be the Kac-Moody algebra given by the Cartan matrix $C$.
This corresponds to adding $N-1$ more nodes to the Dynkin diagram of $V$, each connected to the previous one by a single line.

In the same way as for $U$ and $V$, the generators $e_0$ and $f_0$ give rise to a $\mathbb{Z}$-grading of $W$,
where each subspace 
$W_p$ constitutes a representation ${\bf t}_p$ of $\g$, but also a representation of $\sl_N$.
Considering $V$ as a subalgebra of $W$ we can write the basis 
elements of $W_1$ and $W_{-1}$ as $E_{\cM}{}_a$ and $F^{\cM}{}^b$, respectively, where $a,b,\ldots=0,1,\ldots,N-1$, and
\begin{align}
E_{\cM}{}_{0} &= E_\cM, & E_{\cM}{}_{(-m)} &= [[ \cdots [[e_{m},e_{m+1}],e_{m+2}], \ldots, e_{-1}],E_\cM],\nn\\
F^{\cM}{}^{0} &= F^\cM, & F^{\cM}{}^{(-m)} &= (-1)^m[[ \cdots [[f_{m},f_{m+1}],f_{m+2}], \ldots, f_{-1}],F^\cM].
\end{align}
For $p \geq 2$, the subspace $W_{p}$ is then spanned by the elements
\begin{align}
E_{\cM_1\cdots\cM_p}{}_{\ a_1\cdots a_p} &= \dlb E_{\cM_1}{}_{\,a_1},\dlb E_{\cM_2}{}_{\,a_2},\ldots, 
\dlb E_{\cM_{p-1}}{}_{\,a_{p-1}},E_{\cM_p}{}_{\,a_p} \drb \cdots \drb\drb,
\end{align}
and $W_{-p}$ by the elements
\begin{align}
F^{\cM_1\cdots\cM_p}{}^{\ a_1\cdots a_p} &= \dlb F^{\cM_1}{}^{\,a_1},\dlb F^{\cM_2}{}^{\,a_2},\ldots, 
\dlb F^{\cM_{p-1}}{}^{\,a_{p-1}},F^{\cM_p}{}^{\,a_p} \drb \cdots \drb\drb.
\end{align}
Following the steps in \cite{Palmkvist:2007as} 
it is straightforward to show that the structure constants of the triple system $W_{-1}$ are related to those of
$V_{-1}$ as
\begin{align} \label{simpleformula}
h_\cM\,{}^\cN\,{}_\cP\,{}^\cQ{}_a{}^b{}_c{}^d &= ( [[E_\cM{}_a,F^\cN{}^b],E_\cP{}_c]| F^{\cQ}{}^d ) \nn\\
&= g_\cM{}^\cN{}_\cP{}^\cQ\, \de_a{}^b \de_c{}^d-\de_\cM{}^\cN\de_{\cP}{}^\cQ\,\de_a{}^b \de_c{}^d
+\de_\cM{}^\cN\de_{\cP}{}^\cQ\,\de_c{}^b \de_a{}^d,
\end{align}
and if we antisymmetrize in $a$ and $c$ we obtain
\begin{align} \label{observation}
h_\cM{}^{\cN}{}_{\cP}{}^\cQ{}_{[a}{}^b{}_{c]}{}^d\,
&=(g_\cM{}^\cN{}_\cP{}^\cQ - 2\, \de_\cM{}^\cN\de_{\cP}{}^\cQ)\de_{[a}{}^b \de_{c]}{}^d = 
f_\cM{}^\cN{}_\cP{}^\cQ \,\de_{[ac]}{}^{bd}.
\end{align}
Thus we get back the structure constants ({\ref{lilla-f}}) for the triple system $\bor_{-1}$, times $\de_{[a}{}^b \de_{c]}{}^d$.
As we will see next, this relation between the two triple systems can be viewed as the reason why 
$U$ and $W$ lead to the same $p$-form spectrum, or to be precise, why ${\bf t}_p={\bf r}_p$ for $1 \leq p \leq N$, which is the main result of this paper.

As for $U$, each representation ${\bf t}_p$ is determined by the lower
indices in the tensor
\begin{align} \label{h-innerprod}
h_{\cM_1\cdots\cM_{p}}&{}^{\cN_1\cdots\cN_{p}}{}_{[a_1 \cdots a_p]}{}^{b_1 \cdots b_p}
= (  E_{\cM_1\cdots\cM_p}{}_{\ a_1\cdots a_p}| F^{\cN_1\cdots\cN_p}{}^{\ b_1\cdots b_p}).
\end{align}
In the same way as we obtained (\ref{borcherdsutrakning}) for $U$, we now obtain 
\begin{align}
h_{\cM_1\cdots\cM_{p}}&{}^{\cN_1\cdots\cN_{p}}{}_{[a_1 \cdots a_p]}{}^{b_1 \cdots b_p}=
( E_{\cM_1\cdots\cM_{p}\,[a_1 \cdots a_p]} | F^{\cN_1\cdots\cN_{p}\ b_1 \cdots b_p})\nn\\
&=\sum_{i=1}^{p-1} \sum_{j=i+1}^p
h_{\cM_i}{}^{\cN_1}{}_{\cM_j}{}^\cP{}_{[a_i}{}^{b_1}{}_{a_j}{}^c \times \nn\\
&\qquad\qquad\times h_{\cM_1 \cdots \cP \cdots \cM_p}{}^{\cN_2\cdots\cN_p}
{}_{a_1 \cdots a_{i-1}a_{i+1}\cdots a_{j-1}|c|a_{j+1}\cdots a_p]}{}^{b_2\cdots b_p}\nn\\
&\qquad\qquad\qquad-h_{\cM_p}{}^{\cN_1}{}_{\cM_{p-1}}{}^\cP {}_{[a_p}{}^{b_1}{}_{a_{p-1}}{}^c
h_{\cM_1 \cdots \cM_{p-2}\cP}{}^{\cN_2\cdots\cN_p}{}_{a_1 \cdots a_{p-2}]c}{}^{b_2\cdots b_p}
\end{align}
for $W$, where we have simplified the notation by writing
\begin{align}
\cM_1 \cdots \cP \cdots \cM_p = \cM_1 \cdots \cM_{i-1}\cM_{i+1}\cdots \cM_{j-1}\cP\cM_{j+1}\cdots \cM_p.
\end{align}
The difference compared to (\ref{borcherdsutrakning}) is that $f$ is replaced by $h$,
that each ${\bf r}_1$ index is accompanied by an $\sl_N$ index and, most important, that the prefactor $(-1)^{i+p}$ is replaced by $1$. We will now show, by induction over $p$, that
\begin{align}
h_{\cM_1\cdots\cM_{p}}&{}^{\cN_1\cdots\cN_{p}}{}_{[a_1 \cdots a_p]}{}^{b_1 \cdots b_p}
=(-1)^{\sigma(p)}\de_{[a_1\cdots a_p]}{}^{b_1\cdots b_p}\,f_{\cM_1\cdots\cM_p}{}^{\cN_1\cdots\cN_{p}}, \label{indpast}
\end{align}
for all integers $p \geq 1$, where $\sigma(p)={p(p-1)}/2$.
For $p=1$ we have
\begin{align}
h_\cM{}^\cN{}_a{}^b = (E_\cM{}_a|F^\cN{}^b)= \de_a{}^b \de_\cM{}^\cN = \de_a{}^b \la E_\cM | F^\cN \ra = \de_a{}^b g_\cM{}^\cN.
\end{align}
Assume now that (\ref{indpast}) holds for $p=q-1$, where $q$ is some integer $q\geq 2$. Then 
\begin{align}
h_{\cM_1\cdots\cM_{q}}&{}^{\cN_1\cdots\cN_{q}}{}_{[a_1 \cdots a_q]}{}^{\ b_1 \cdots b_q}=
\la E_{\cM_1\cdots\cM_{q}\,[a_1 \cdots a_q]} | F^{\cN_1\cdots\cN_{q}\,b_1 \cdots b_q}\ra\nn\\
&=\sum_{i=1}^{q-1} \sum_{j=i+1}^q
h_{\cM_i}{}^{\cN_1}{}_{\cM_j}{}^\cP{}_{[a_i}{}^{b_1}{}_{a_j}{}^c \times \nn\\
&\qquad\qquad\times h_{\cM_1 \cdots \cP \cdots \cM_q}{}^{\cN_2\cdots\cN_q}
{}_{a_1 \cdots a_{i-1}a_{i+1}\cdots a_{j-1}|c|a_{j+1}\cdots a_q]}{}^{b_2\cdots b_q}\nn\\
&\qquad-h_{\cM_q}{}^{\cN_1}{}_{\cM_{q-1}}{}^\cP {}_{[a_q}{}^{b_1}{}_{a_{q-1}}{}^c
h_{\cM_1 \cdots \cM_{q-2}\cP}{}^{\cN_2\cdots\cN_q}{}_{a_1 \cdots a_{q-2}]c}{}^{b_2\cdots b_q}\nn\\
&=\sum_{i=1}^{q-1} \sum_{j=i+1}^q
f_{\cM_i}{}^{\cN_1}{}_{\cM_j}{}^\cP\de_{[a_i a_j}{}^{b_1c} \times \nn\\
&\qquad\qquad\times (-1)^{\sigma(q-1)} f_{\cM_1 \cdots \cP\cdots \cM_q}{}^{\cN_2\cdots\cN_q}
\de_{a_1 \cdots a_{i-1}a_{i+1}\cdots a_{j-1}|c|a_{j+1}\cdots a_q]}{}^{[b_2\cdots b_q]}\nn\\
&\qquad-(-1)^{\sigma(q-1)} f_{\cM_q}{}^{\cN_1}{}_{\cM_{q-1}}{}^\cP \de_{[a_q a_{q-1}}{}^{b_1 c}
f_{\cM_1 \cdots \cM_{q-2}\cP}{}^{\cN_2\cdots\cN_q}\de_{a_1 \cdots a_{q-2}]c}{}^{[b_2\cdots b_q]}\nn\\
&=\sum_{i=1}^{q-1} \sum_{j=i+1}^q
(-1)^{\sigma(q-1)} f_{\cM_i}{}^{\cN_1}{}_{\cM_j}{}^\cP f_{\cM_1 \cdots \cP\cdots \cM_q}{}^{\cN_2\cdots\cN_q}
\de_{[a_i a_1 \cdots a_{i-1}a_{i+1}\cdots a_q]}{}^{b_1b_2\cdots b_q}\nn\\
&\qquad-(-1)^{\sigma(q-1)} f_{\cM_q}{}^{\cN_1}{}_{\cM_{q-1}}{}^\cP 
f_{\cM_1 \cdots \cM_{q-2}\cP}{}^{\cN_2\cdots\cN_q}\de_{[a_q a_1 \cdots a_{q-1}]}{}^{b_1 b_2\cdots b_q}\nn\\
&=\sum_{i=1}^{q-1} \sum_{j=i+1}^q
(-1)^{{\sigma(q-1)}+i+1} f_{\cM_i}{}^{\cN_1}{}_{\cM_j}{}^\cP f_{\cM_1 \cdots \cP\cdots \cM_q}{}^{\cN_2\cdots\cN_q}
\de_{[a_1 \cdots a_q]}{}^{b_1\cdots b_q}\nn\\
&\qquad+ (-1)^{{\sigma(q-1)}+q} f_{\cM_q}{}^{\cN_1}{}_{\cM_{q-1}}{}^\cP 
f_{\cM_1 \cdots \cM_{q-2}\cP}{}^{\cN_2\cdots\cN_q}\de_{[a_1 \cdots a_{q}]}{}^{b_1\cdots b_q}\nn\\
&=(-1)^{{\sigma(q-1)+q-1}}\de_{[a_1 \cdots a_{q}]}{}^{b_1\cdots b_q}f_{\cM_1\cdots\cM_{q}}{}^{\cN_1\cdots\cN_{q}}\nn\\
&=(-1)^{{\sigma(q)}}\de_{[a_1 \cdots a_{q}]}{}^{b_1\cdots b_q}f_{\cM_1\cdots\cM_{q}}{}^{\cN_1\cdots\cN_{q}},
\label{borcherdsutrakning2}
\end{align}
where we first have inserted the assumption of the induction, and then used (\ref{borcherdsutrakning}).
By the principle of induction, it follows that
(\ref{indpast}) holds for all integers $p \geq 1$. Since the lower ${\bf r}_1$ indices on the left hand side of  
(\ref{indpast}) determine ${\bf r}_p$, and those on the right hand side determine ${\bf t}_p$,
we conclude that ${\bf r}_p={\bf t}_p$ as long as the delta factor does not vanish, that is, for $1 \leq p \leq N$.

\subsubsection*{Acknowledgments}
I would like to thank Jesper Greitz and Paul Howe for sharing a draft of \cite{Greitz:2012}, and
Jesper Greitz for discussions.
The work is supported by IISN -- Belgium (conventions 4.4511.06 and 4.4514.08), by the Belgian Federal Science Policy Office 
through the Interuniversity Attraction Pole P6/11.

\bibliographystyle{utphysmod2}


\providecommand{\href}[2]{#2}\begingroup\raggedright\endgroup

\end{document}